\documentclass[twocolumn,superscriptaddress,showpacs,preprintnumbers,amsmath,amssymb,notitlepage,longbibliography]{revtex4-2}

\newcommand{\schro}{Schr\"{o}dinger }
\usepackage{bm}
\usepackage{graphicx}
\usepackage{color} %for change tracking
\usepackage{setspace}
\usepackage[T1]{fontenc}
\usepackage{subfigure}
\begin{document}
\title{Wave manipulation via delay-engineered periodic potentials}

\author{Alessandro Alberucci}
\affiliation{Friedrich Schiller University Jena, Institute of Applied Physics, Abbe Center of Photonics,  Albert-Einstein-Str. 15, 07745 Jena, Germany}
\author{Chandroth P. Jisha}
\affiliation{Friedrich Schiller University Jena, Institute of Applied Physics, Abbe Center of Photonics, Albert-Einstein-Str. 15, 07745 Jena, Germany}
\author{Monika Monika}
\affiliation{Friedrich Schiller University Jena, Institute of Solid State Physics and Optics, Abbe Center of Photonics, Max-Wien-Platz 1, 07743 Jena, Germany}
\author{Ulf Peschel}
\affiliation{Friedrich Schiller University Jena, Institute of Solid State Physics and Optics, Abbe Center of Photonics, Max-Wien-Platz 1, 07743 Jena, Germany}
\author{Stefan Nolte}
\affiliation{Friedrich Schiller University Jena, Institute of Applied Physics, Abbe Center of Photonics, Albert-Einstein-Str. 15, 07745 Jena, Germany}
\affiliation{Fraunhofer Institute for Applied Optics and Precision Engineering, Albert-Einstein-Straße 7, 07745 Jena, Germany}

\date{\today}% It is always \today, today,
             %  but any date may be explicitly specified

\begin{abstract}
 We discuss the semi-classical transverse trapping of waves by means of an inhomogeneous gauge field. In the proposed scheme a temporally-periodic perturbation is shifted in time, the imparted delay being dependent on the transverse direction. We show that, due to the Kapitza effect, an effective potential proportional to the square of the transverse derivative of the delay arises. On a more physical ground, the delay induces a transversely-varying periodic force acting on the wave, in turn providing a phase delay owing to the local modulation of the kinetic energy. Our results are quite generic and can find application in several fields, ranging from cold atoms to optics: accordingly, an experimental proof-of-principle is provided using an optical set-up based upon fiber loops.  
\end{abstract}
%\pacs{Valid PACS appear here}% PACS, the Physics and Astronomy
                             % Classification Scheme.
%\keywords{Suggested keywords}%Use showkeys class option if keyword
                              %display desired

\maketitle

%\tableofcontents

\section{Introduction: gauge fields in time-shifted potentials}

Gauge theories are a fundamental pillar in physical sciences. In classical electromagnetism, different gauges manifest themselves as different values for the scalar $V$ and vectorial potential $\bm{A}$, although the physically-measurable electromagnetic field remains unvaried, the latter being commonly referred to as gauge invariance \cite{Jackson:2001}. Historically, the concept of gauge invariance has been explicitly formalized by Weyl while searching for a unified theory accounting for both general relativity and electromagnetism \cite{Weyl:1929}. The concept of gauge is strictly related to local symmetries: for example, in quantum electrodynamics the electromagnetic field emerges as the correct interaction between electrons by imposing invariance with respect to an arbitrary spatio-temporal phase delay, the latter transformation corresponding to the Lie group U(1). This idea has been then extended by Yang-Mills to different symmetry groups including the non-Abelian case, in turn permitting the elaboration of the Standard Model and its components, such as the quantum chromodynamics.\\
In the last few decades, a great deal of attention has been directed to the generation of gauge fields for neutral particles, including cold atoms and photons \cite{Lin:2009,Struck:2012,Hauke:2012,Anderson:2013,Aidelsburger:2015,Ray:2014,Aidelsburger:2018,Hafezi:2011,Fang2012,Rechtsman:2013_1,Liu:2015,Schine:2016,Rechcinska:2019,Pankov:2019,Wang:2020,Chalabi:2020,Cohen:2020,Bandres:2022}, but also including mechanical systems \cite{Abbaszadeh:2017} and polariton BEC \cite{Tercas:2014,Jamadi:2020}. Stated in simpler terms, the idea is to find effective interactions capable to mimick the action of a gauge field, emblematically represented by the presence of a magnetic field. In fact, systems subject to a magnetic-like interaction present a plethora of exotic effects with various applications, including the quantum spin Hall effect and related topological phenomena \cite{Thouless:1982,Hafezi:2011,Tokura:2019}. As is well known, a magnetic field acts only on charged particles. To overcome this intrinsic limitation, some tricks need to be used, often involving the modulation of the system in time \cite{Wilczek:1984,Fang2013,Anderson:2013}. Indeed, several works showed how periodically-driven systems (the so called Floquet systems \cite{Shirley1965,Yin:2022}) can be tailored to achieve time-averaged Hamiltonians featuring topological effects \cite{Goldman:2014,Eckardt:2015,Wen:2020,Nuske:2020}. \\
In this Article we discuss the propagation of waves in a potential [dubbed $V(x,t)$] periodic in time $t$ [i.e., $V(x,t)=V(x,t+T)$] and inhomogeneous across the spatial dimension(s) [$\partial V(x,t)/\partial x \neq 0$]. The idea is to control the wave propagation by imposing a space-dependent temporal delay [local transformation $t\rightarrow t -\tau(x)$] on the periodic potential, that is, to introduce a point-dependent gauge transformation corresponding to a shift in time. Our idea closely resembles and indeed generalizes the theoretical proposal for optical waveguiding based upon an inhomogeneous gauge field due to dispersion shifting \cite{Lin2014a}, recently experimentally demonstrated in a waveguide array \cite{Lumer:2019}. The framework we are introducing here also provides a deeper understanding for the existence of waveguides based upon geometric phase in inhomogeneously rotated liquid crystals \cite{Slussarenko:2016,Jisha2017}. \\
To focus on real physical systems, we use the \schro equation, the latter apt to describe both the evolution in time of massive particles and the evolution in space of photons in the monochromatic regime \cite{Longhi:2009}. For homogeneous distribution of the delay [$V(x,t)=U(x)f(t)$], it has been already shown in several papers that, even if the average potential is vanishing, an effective potential proportional to the square of the space derivative of the potential [i.e., $\propto \left( d U/d x\right)^2$] arises due to the local modulation of the kinetic energy, an effect named after its discoverer Kapitza \cite{Kapitza1951,Cook1985,Rahav2003,Longhi2011,Muniz:2019_1}. Here we generalize the Kapitza model to the presence of an inhomogeneous delay, showing that a straightforward generalization of the effective time-independent potential predicts the presence of a new term dependent on the spatial derivative of the local phase shift [i.e., $\propto \left(d \tau/ dx \right)^2$]. We confirm the accuracy and the limit of applicability of this effective model by using numerical simulations of the time-dependent \schro equation in one transverse dimension. We also show how the different contributions to the Kapitza potential interact with each other, in particular how parity strongly affects the wave behavior. Our theoretical results are experimentally verified in fiber loops, where a discrete set of optical pulses follows the \schro equation with very good accuracy \cite{Muniz:2019_1}.  \\
The Paper is organized as follows. In Sec.~\ref{sec:geometry} the type of time-periodic potential used in this paper is depicted. In Sec.~\ref{sec:Kapitza_model} we show how the action of the transverse-dependent delay can be modelled like a time-independent potential by adapting the Kapitza approach to this specific case. In Sec.~\ref{sec:pot_even_delay_even} the transverse confinement of light is theoretically demonstrated for the simplest case of an even delay and flat potential. In Sec.~\ref{sec:odd_potential} we discuss the wave confinement when the potential is odd symmetric, whereas the delay is kept even. In Sec.~\ref{sec:odd_delay} we prove that an odd profile for the delay can be designed to provide full confinement of the wave in a given spatial region. In Sec.~\ref{sec:fiber_loop} an experimental verification of the wave confinement due to the inhomogeneous delay is reported. In Sec.~\ref{sec:discussion} we discuss the potential impact of our results, providing a short survey of the physical systems where our results can be of interest. Sec.~\ref{sec:conclusions} contains a compact summary of our work and its potential role in future research.

\section{Wave propagation in the presence of non-uniform delay}
\label{sec:geometry}
We suppose that the complex wave $\psi$ obeys a standard \schro (1+1)D equation
\begin{equation} \label{eq:SE}
 i\hbar \frac{\partial \psi}{\partial t}= -D \frac{\partial^2 \psi}{\partial x^2} + V(x,t)\psi,
\end{equation}
where $D=\hbar^2/(2m)$ in the case of quantum particles of mass $m$. The correspondence with the optical case is described in Appendix~\ref{sec:optical_case}. Variable $x$ (spatial dimension both for matter and optical waves) and $t$ (time for matter, propagation distance for optical waves) correspond to the \textit{transverse} and the \textit{longitudinal} coordinates, respectively. We suppose the external semi-classical potential to be periodic with period $T$ by assuming $V(x,t)=V(x,t+T)$. 
In terms of Fourier series, it is $V(x,t)=\sum_n V_n(x)e^{-2\pi i n \frac{t}{T}}$. In agreement with the main purpose of this paper, hereafter the average potential is set to vanish, yielding $V_0(x)=0$. The existence of steady states in the presence of a periodic perturbation superposed to a time-independent bounding potential has been formally addressed in Ref.~\cite{Sambe:1973}. \\
\begin{figure}
\includegraphics[width=0.45\textwidth]{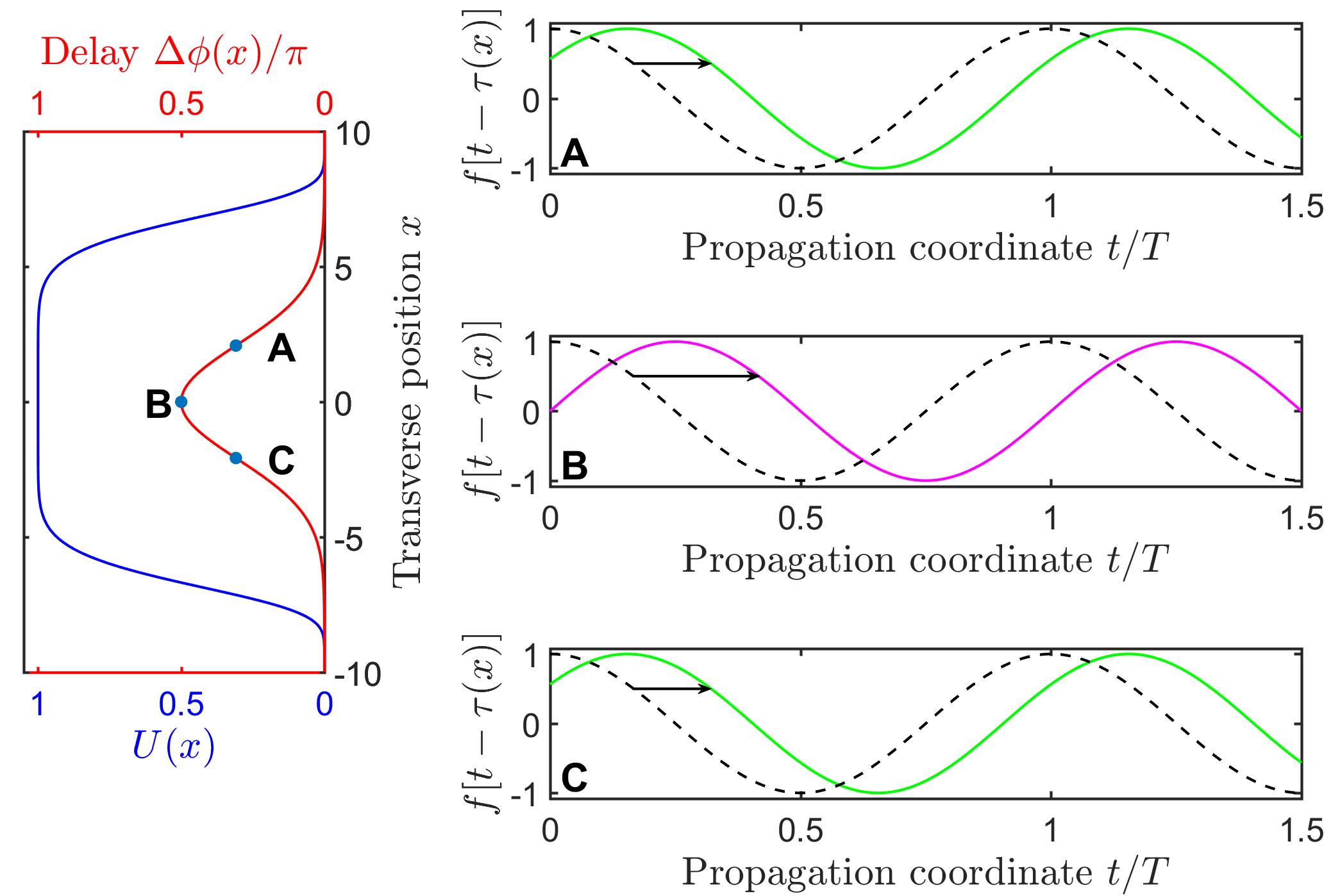}
\caption{\label{fig:sketch} {Left side: Local amplitude of the oscillating field $U(x)$ (blue curve) and of the local delay $\phi(x)/\pi=2\tau(x)/T$ (red curve) versus the transverse coordinate $x$. Right side: periodic modulation of the potential along the normalized propagation coordinate $t/T$ at three different values of the transverse position $x$. The black dashed line is the longitudinal modulation $f(t)$ (defined in Sec.~\ref{sec:Kapitza_model}) for $\tau=0$. The curves are plotted for the special case $f(t)=\cos(2\pi t/T)$, $U(x)=e^{-x^8/w^8}$ and $\tau(x)=\frac{\pi}{2} e^{-x^2/w^2_\mathrm{delay}}$. }}
\end{figure}
We are interested in the design of a potential capable of supporting transverse confinement by inducing an $x$-dependent gauge field $\bm{A}$. Our core idea is to consider an $x$-dependent temporal delay $\tau(x)$ by setting (see Fig.~\ref{fig:sketch})
\begin{equation} \label{eq:delayed_potential}
  V(x,t)=U(x)f\left[t-\tau(x) \right],
\end{equation}
the latter corresponding in the Fourier domain to $V_n(x)= U(x) f_n e^{-2 \pi i n \frac{\tau(x)}{T}}$, where $f_n$ are the Fourier coefficients for $\tau(x)=0$, $f(t)=\sum_n f_n e^{-2\pi i n \frac{t}{T}}$. %The quantity $V^{(0)}(x)$ thus represents the transverse profile of the potential in the absence of any delay. 
The important physical point is that now the spatial and temporal dependence of the external potential cannot be  factorized. Application of the gauge transformation provided by the indefinite integral $\psi=\varphi e^{-\frac{i}{\hbar}\int{V(x,t)dt}}$ to Eq.~\eqref{eq:SE} shows that $\varphi$ evolves according to \cite{Lin2014a}
\begin{equation}
     i\hbar \frac{\partial \varphi}{\partial t}= \frac{1}{2m}\left( -i\hbar \frac{\partial }{\partial x} - A_x \right)^2\varphi,
\end{equation}
where we introduced the effective potential vector
\begin{equation} \label{eq:potential_vector}
  A_x= %\frac{1}{\kappa}
	\int{ \left(\frac{\partial U}{\partial x}f - U  \frac{d f}{{d t}}   \frac{d\tau}{dx} \right) dt}.
\end{equation} 
Interpreted in terms of synthetic electromagnetic interactions, Eq.~\eqref{eq:potential_vector} demonstrates that the wave behaves like a charged particle under the action of an effective electromagnetic field, the magnetic component being null and the electric field being expressed by %$\bm{E}_\mathrm{eff}=\frac{\hat{x}}{\kappa}\left(\partial_x V^{(0)}  + \tau^\prime \partial_t V^{(0)}\right)$ with $\tau^\prime=d\tau/dx$. 
 $\bm{E}_\mathrm{eff}= -\left[f\partial_x U  -  U (df/d\tau) (d\tau/dx) \right]\hat{x}$, where %$\tau^\prime=d\tau/dx$ and 
 $\hat{x}$ is the unit vector along the $x$ direction \cite{Chalabi:2020,Derrico:2021}. In fact, we find the same expression for $\bm{E}_\mathrm{eff}$ by computing directly the spatial gradient of Eq.~\eqref{eq:delayed_potential}, confirming the invariance of the synthetic electromagnetic field with respect to gauge transformations.  
When the delay of the potential is constant across $x$, Eq.~\eqref{eq:potential_vector} confirms that the Kapitza potential stems from the modulation of the kinetic momentum.  Even when $\partial_x U=0$, $\bm{E}_\mathrm{eff}$ is not vanishing: in fact, it is proportional to the transverse derivative of the delay $\tau(x)$. The physical interpretation is straightforward: even if the amplitude of the oscillating potential is constant, the inhomogeneous delay generates locally a gradient in the potential energy, the latter in turn modulating the local momentum, as in the original Kapitza effect.

\section{Effective time-independent potential}
\label{sec:Kapitza_model}
We first consider the family of real and factorizable potentials $V(x,t)=f(t)U(x)$. In this case, Kapitza \cite{Kapitza1951} was the first to show that the $x$-dependent modulation of the kinetic energy yields an effective potential \cite{Rahav2003}
\begin{equation} \label{eq:Veff}
V_\mathrm{eff}= \frac{T^2}{2m}\sum_{n\neq 0}\frac{1}{4\pi^2 n^2}\left|\frac{\partial V_n}{\partial x}\right|^2 = \frac{T^2}{2m}\sum_{n\neq 0}\frac{|f_n|^2}{4\pi^2 n^2}\left(\frac{\partial U}{\partial x}\right)^2.
\end{equation}
It has been proven theoretically \cite{Cook1985,Rahav2003} and experimentally \cite{Muniz:2019_1} that this class of potentials can induce transverse trapping of waves. \\
Next step is to find another gauge transformation capable to enlighten how the modulation of the kinetic energy impacts the wave profile \cite{Reiss:2008,Reiss:2017}. 
Given that in the presence of non-uniform delay $\partial_x V_n=f_n\left[\partial_x U - \left(2\pi i n /T \right) U  \partial_x \tau \right]$, Eq.~\eqref{eq:Veff} can be recast as
%In the presence of a delay $\tau(x)$, $V_n(x)=V^{}$.
% After setting $\psi=\varphi \exp{\left(-\frac{i}{\hbar} \sum_{n=0}^\infty \int{E_n(x,t) dt} \right)}$ (see Ref.~\cite{Alberucci:2020} for details), the Kapitza potential at the order $O(T^2)$ reads
\begin{equation}
   V_\mathrm{eff}= \frac{T^2}{8 \pi^2 m}\sum_{n\neq 0} |f_n|^2 \left[\frac{1}{ n^2}\left(\frac{\partial U}{\partial x}\right)^2 + U^2 \left( \frac{\partial \phi}{\partial x}\right)^2   \right], \label{eq:Veff_delay}
\end{equation} 
where we defined $\phi (x)=2\pi \tau(x)/T$.
 Eq.~\eqref{eq:Veff_delay} accurately models the wave propagation in a periodic potential, if the longitudinal period $T$ is not too long with respect to the dispersion length of the wave (see Appendix~\ref{sec:model_accuracy} for further details). The effective potential now depends both on the gradient of the oscillation amplitude and on the gradient of the delay. The main difference is the absence of the term $n^{-2}$ in the term depending on the delay, thus making the contribution of higher harmonics more important in the latter case.\\
Hereafter, we will focus on the sinusoidal case $f(t)=\cos\left[2\pi \left(t-\tau(x) \right)/T+\varphi_\mathrm{in}\right]$ to simplify the math, despite the physical phenomena of interest remains substantially untouched when more complicated periodic functions are assumed. Quantity $\varphi_\mathrm{in}$ is a uniform (along $x$) initial phase delay, which correspond physically to a shift of the temporal axis.
The quasi-mode is then given by (see Ref.~\cite{Alberucci:2020} and Appendix~\ref{sec:model_accuracy})
\begin{equation}  \label{eq:quasimode}
  \psi(x,t)\approx g(x)e^{-i\frac{U(x)T}{2\pi\hbar}\sin\left[\frac{2\pi t}{T} +\varphi_\mathrm{in} - \phi(x)\right]} e^{-i\frac{E_0 t}{\hbar}},
\end{equation}
where $g(x)$ is the eigenfunction of a time-invariant \schro equation with the potential given by Eq.~\eqref{eq:Veff_delay}, and $E_0$ the corresponding eigenvalue. Whereas for vanishing transverse gradient of the delay (i.e., $\partial \phi/\partial x=0$) there are instants where the wavefront is planar, in the presence of delay the wavefront of the quasi-mode is never planar, even at the lowest order of approximation in the modulation period $T$. \\
\begin{figure}
\includegraphics[width=0.49\textwidth]{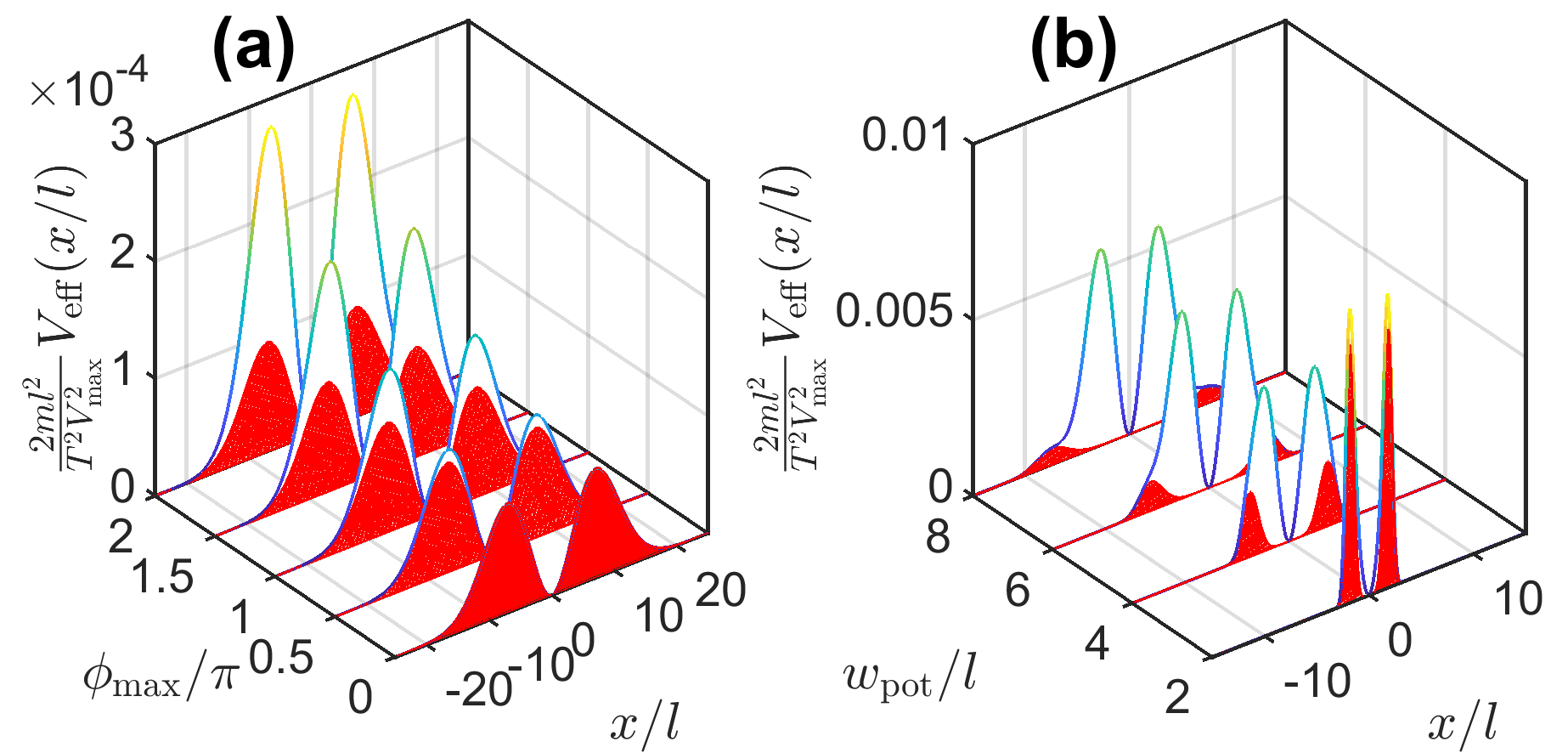}
\caption{\label{fig:potential} {The normalized effective potential $\frac{2ml^2}{T^2V^2_\mathrm{max}}V_\mathrm{eff}(x/l)$ versus the normalized transverse coordinate $x/l$ (a) as the phase delay contribution is increased for a fixed $U(x/l)$, and (b) as the contribution from $U(x/l)$ is decreased for a fixed delay distribution $\phi(x/l)$. The red areas represent the effective potential for $\phi=0$. In (a) $U(x)=V_\mathrm{max}e^{-x^2/w^2_\mathrm{pot}}$ with $w_\mathrm{pot}/l=10$, and $ \phi= \phi_\mathrm{max}e^{-x^2/w^2_\mathrm{\phi}}$ with $w_\mathrm{\phi}/l$=20; in  (b) $\phi$ has the same functional form but with $\phi_\mathrm{max}=\pi$ and $w_\mathrm{\phi}/l$=4, whereas $U(x)=V_\mathrm{max}e^{-x^N/w^N_\mathrm{pot}}$ with $N=4$. In both the panels we fixed $f(t)=\cos\left(2\pi t/T \right)$.}}
\end{figure}
\begin{figure*}
\includegraphics[width=0.95\textwidth]{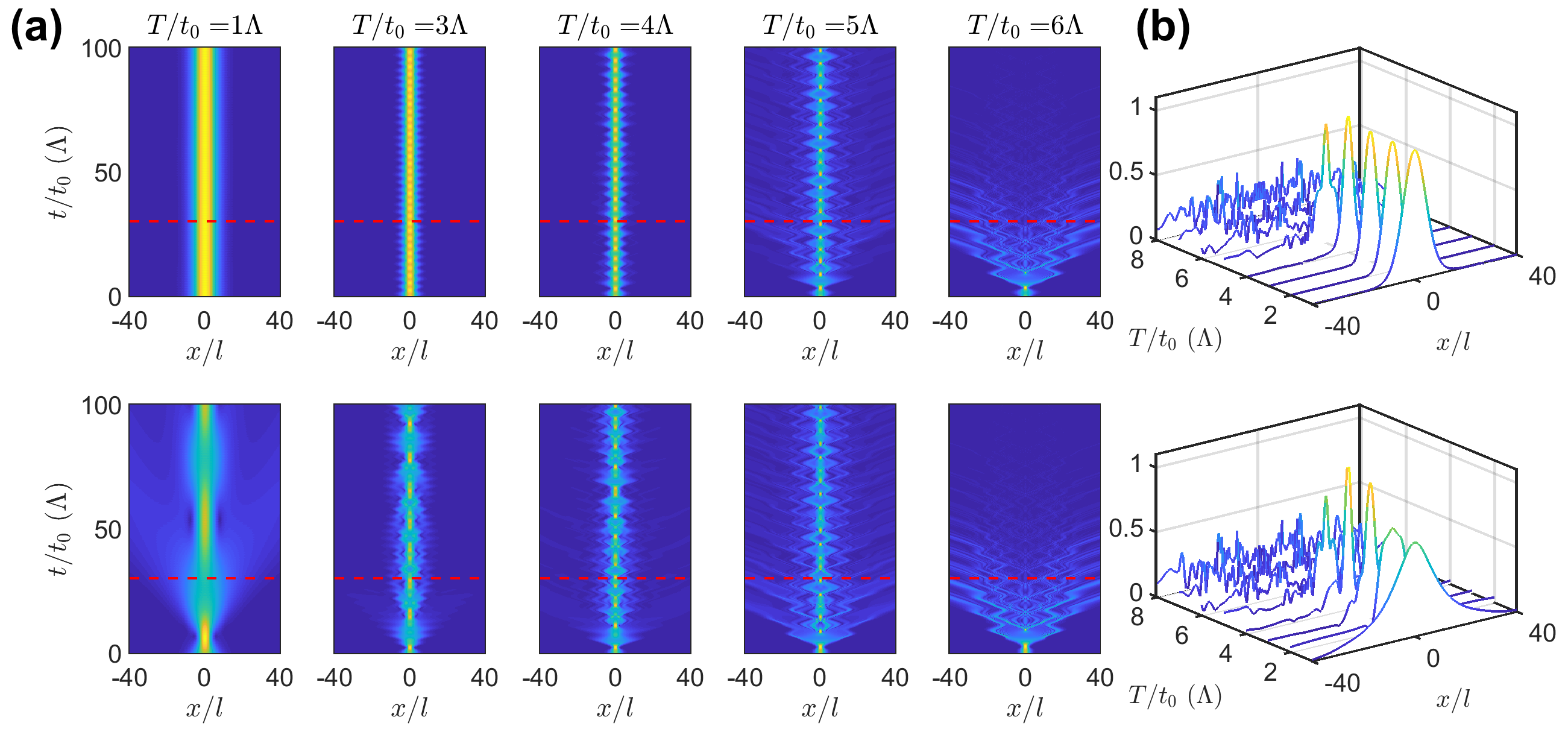}
\caption{\label{fig:numerics_gaussian} {Propagation of the quasi-mode given by Eq.~\eqref{eq:quasimode} in a delayed  potential featuring $U(x)=0.5e^{-x^4/w_\mathrm{pot}^4}$  and $\phi(x)=\pi w^{-x^2/w^2_\phi}$, with $w_\mathrm{pot}/l=200$ and $w_\phi/l=20$. In (a) $|\psi|^2$ on the propagation plane is plotted for increasing period $T/t_0$ from left to right, when in the input condition the phase front curvature is accounted for (top row) and when a planar wavefront is assumed (bottom row). In (b) the modulus of the wave $|\psi|$ calculated in $t/t_0=30~\Lambda$ [section corresponding to red dashed lines in (a)] is plotted versus $x/l$ for periods $T/t_0$ ranging from $\Lambda$ to 6$\Lambda$. Here $\varphi_\mathrm{in}=0$.}}
\end{figure*}
In Fig.~\ref{fig:potential} the general behavior of Eq.~\eqref{eq:Veff_delay}, driven by the interplay between the $x$-derivatives of $U$ and $\tau$, is described. Hereafter our results will be presented versus the transverse normalized coordinate $x/l$, where $l$ is an arbitrary length (see Appendix~\ref{sec:normalization} for more details). From Eq.~\eqref{eq:Veff_delay}, the effective potential $V_\mathrm{eff}$ then scales  with the inverse of $l^2$. If the same functional form for the two quantities is chosen [Fig.~\ref{fig:potential}(a)], no strong deformations of the potential are observed when the maximum delay $\phi_\mathrm{max}$ is increased. Clearly, the term stemming from the delay becomes dominant for large enough delays. On the other side, the contribution from $\partial_x U$ can be made negligible by choosing a flat-top profile for $U(x)$, see Fig.~\ref{fig:potential}(b). In fact, Eq.~\eqref{eq:Veff_delay} then provides 
\begin{equation} \label{eq:Veff_only_delay}
V_\mathrm{eff}\approx \frac{T^2 V_\mathrm{cen}^2}{16 \pi^2 m} \left( \frac{\partial \phi}{\partial x}\right)^2,
\end{equation}
 where $V_\mathrm{cen}=U(x=0)$ is the amplitude of the oscillation in correspondence to the time-shifted zone. Thus, for a given $\tau(x)$, the height of the effective potential can be increased by using larger amplitude oscillations.\\
In the following sections we will concentrate on numerical simulations of the \schro equation to demonstrate the relationship between the local delay and the effective potential \eqref{eq:Veff}. For the sake of concreteness, we will take $D=1.27\times 10^{-14}~$Jm$^2$ and $l=1\mu$m \footnote{This value corresponds to the optical case discussed in the Appendix~\ref{sec:optical_case} for the dispersion parameter in Eq.~\eqref{eq:SE}}. The propagation coordinate is then normalized with respect to the time $t_0=\hbar l^2/D$. To get rounded numbers for the propagation coordinate in the simulations (originally run in the optical domain), we also define the normalized time $\Lambda=0.8$ (in the real time thus corresponding to $T=0.8t_0$), corresponding to a length of 10~$\mu$m for the optical values described in Appendix~\ref{sec:optical_case}. The generalization to other dispersion values can be found by using the normalized equation discussed in Appendix~\ref{sec:normalization}.  
\\
%\textcolor{red}{CALCULATION OF THE NORMALIZATION: $2n_0 k_0^2 l^2 \Delta n=\frac{V l^2}{D}$. To have $V=\Delta n$ (simulations: 1micron wavelength and n0=1) follows $D=\frac{1}{2n_0 k_0^2}$. Then the normalized propagation coordinate is $t/t_0$ with $t_0=\hbar l^2/D$. Also $2k_0n_0 l^2=\hbar l^2/D$ providing $2k_0n_0=\hbar/D \rightarrow \hbar=1/k_0$. Then in the optical simulations $t_0=\frac{1}{k_0 }l^2 2n_0 k_0^2=2n_0k_0l^2=2\times 1 \times \frac{2\pi}{10^-6} \times 10^{-12}=4\pi\times 10^{-6}=\approx 12.5\mu m$.}

\section{Propagation in the case of a Gaussian-distributed delay}
\label{sec:pot_even_delay_even}
Given we want to focus on the role played by $\tau(x)$, let us suppose that the oscillation amplitude of the periodic potential is constant along $x$, thus yielding $\partial_x U=0$ % and $A_x=(1/\kappa)\tau^\prime f(t)U(x)$. 
 and $A_x=-(d\tau/dx) U(x) f(t)$. Figure~\ref{fig:numerics_gaussian} depicts the numerical solution of Eq.~\eqref{eq:SE} when the phase delay is Gaussian shaped. Consequently, the effective potential is M-shaped, as in Fig.~\ref{fig:potential}, thus supporting only leaky modes, i.e., the eigenvalues are imaginary \cite{Rahav2003,Alberucci2013}. As input condition in $t=0$, we took the quasi-mode expression provided by Eq.~\eqref{eq:quasimode}. As predicted, the wave undergoes transverse confinement when the quasi-mode is spatially bounded to the central lobe of the effective potential. When this condition is not satisfied, a significant portion of the wave is eventually coupled to the external regions, where the potential is vanishing as in the center \cite{Alberucci2013}. In agreement with the theory of the Kapitza effect, the mode gets narrower as the modulation period is increased, but at the same time the oscillating components of the quasi-mode get larger and larger. Thus, a shape-preserving wave is observed for short periods (see e.g. $T/t_0=\Lambda$), with increasing periodic oscillations as the period $T/t_0$ gets larger \cite{Alberucci2013}. For $T/t_0=5\Lambda$ higher order components become relevant, and transverse trapping is almost lost, with appreciable radiation losses at each negative half-period. The wave is then fully delocalized for $T/t_0=6\Lambda$. Additionally, the simulations (bottom row in Fig.~\ref{fig:numerics_gaussian}) confirm the fundamental role played by the wavefront in accordance with Eq.~\eqref{eq:quasimode}: when the phase term is not accounted for at the initial time, the confinement is severely hindered. Noticeably, the detrimental effect is stronger at the shortest period, given that in this condition the mode is wider across $x$, thus overlapping more with the phase variations given by $\phi(x)$. 

\section{Even-symmetric delay for an odd instantaneous potential}
\label{sec:odd_potential}

\begin{figure}
\includegraphics[width=0.45\textwidth]{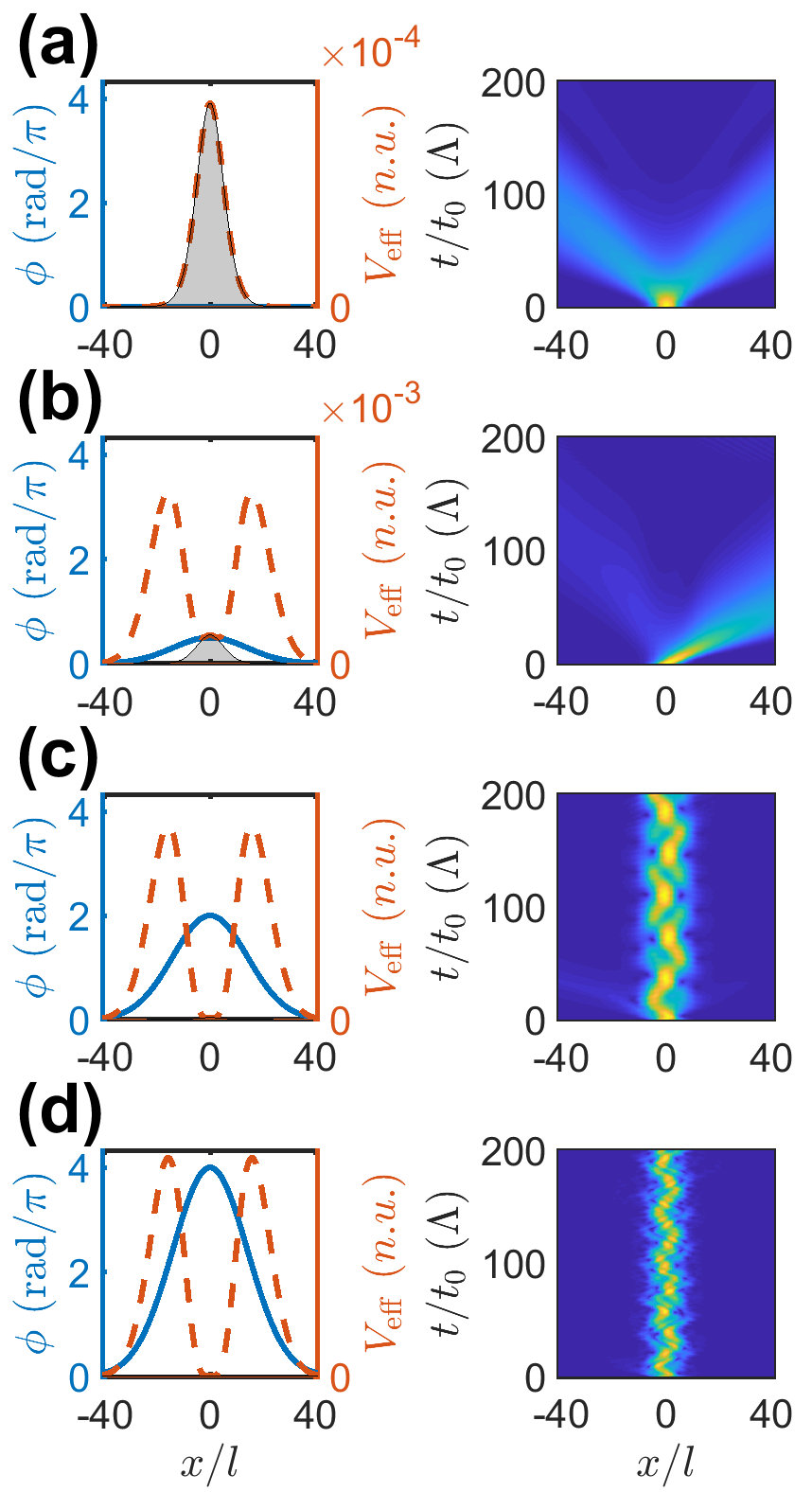}
\caption{\label{fig:odd_potential_vs_max_phase} {Propagation for a Gaussian phase delay $\phi(x)$ of normalized width $20$ and for an odd-symmetric potential $U(x)=0.5 \tanh(x/w_\mathrm{well})$ when the maximum phase delay $\phi_\mathrm{max}$ is 0 (a), $\pi/2$ (b), $2\pi$ (c), and $4\pi$ (d). On the left column blue solid lines correspond to the delay distribution (left axis), whereas dashed red lines and shaded gray area correspond to the effective potential with and without the delay contribution, respectively (right axis). In the right column the intensity distribution on the plane $xt$ is plotted. Here $\varphi_\mathrm{in}=0$.}}
\end{figure}

\begin{figure}
\includegraphics[width=0.45\textwidth]{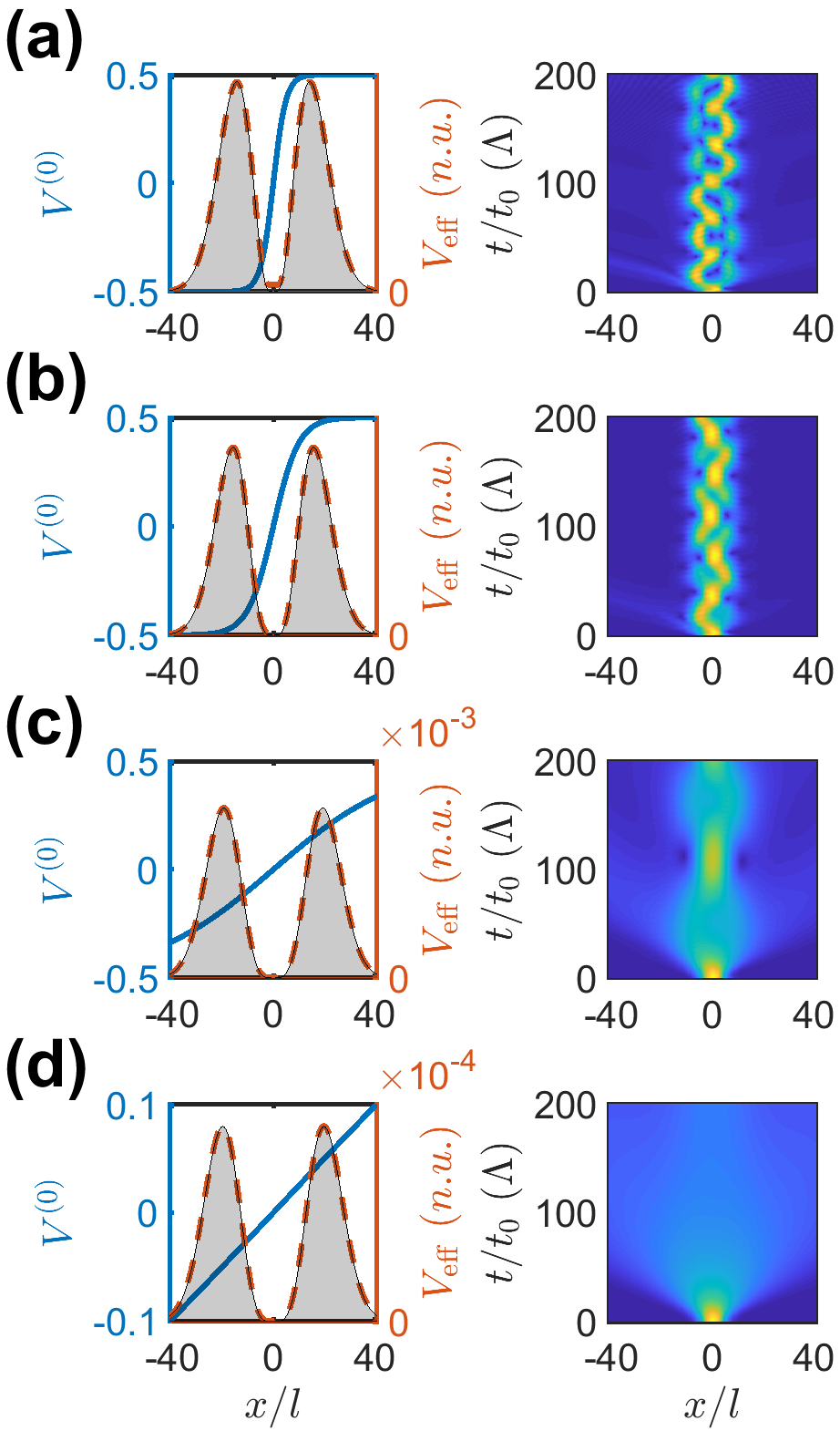} 
\caption{\label{fig:odd_potential_vs_wwell} {Propagation for a Gaussian phase delay $\phi(x)$ of normalized width $20$ and maximum phase $\phi_\mathrm{max}=2\pi$. The underlying potential is $U(x)=0.5 \tanh(x/w_\mathrm{well})$ with $w_\mathrm{well}/l$ equal to 5 (a), 10 (b), 50 (c), and 200 (d). On the left column blue solid lines correspond to $U(x)$ (left axis), whereas dashed red lines and shaded gray area correspond to the effective potential with and without the delay contribution, respectively (right axis). In the right column the intensity distribution on the plane $xt$ is plotted. Here $\varphi_\mathrm{in}=0$.}}
\end{figure}

The overall effective potential given by Eq.~\eqref{eq:Veff_delay} yields a plethora of exotic and counter-intuitive behavior for the wave propagation when the parity of the potential $U(x)$ and $\phi(x)$ is changed. In this section we will start with the case of a Gaussian-shaped distribution for the delay $\tau(x)$ as in Sec.~\ref{sec:pot_even_delay_even}, but the shape of the potential $U(x)$ will be taken odd-symmetric with respect to $x$; specifically, we pick up a hyperbolic tangent shape 
\begin{equation}
U= 0.5 \tanh\left(\frac{x}{w_\mathrm{well}}\right).
\end{equation}
To address the differences with respect to the absence of any potential (stated otherwise, how the periodic potential counteracts the diffractive/dispersive spreading), in the reminder of this section and in the following ones we will take a Gaussian-shaped input 
\begin{equation} \label{eq:gaussian_input}
  \psi(x,t=0)=\psi_0 e^{-\frac{x^2}{w_\mathrm{G}^2}}
\end{equation}
with normalized width $w_\mathrm{G}/l=5$ and flat phase profile in every shown simulation. For reference, the corresponding Rayleigh distance for a free particle is $\approx 8~\Lambda$. Hereafter, we will also fix the modulation period $T/t_0$ equal to $\Lambda$, where we know from the previous discussion that the effective potential model is more accurate than for longer periods.

\begin{figure*}
\includegraphics[width=0.95\textwidth]{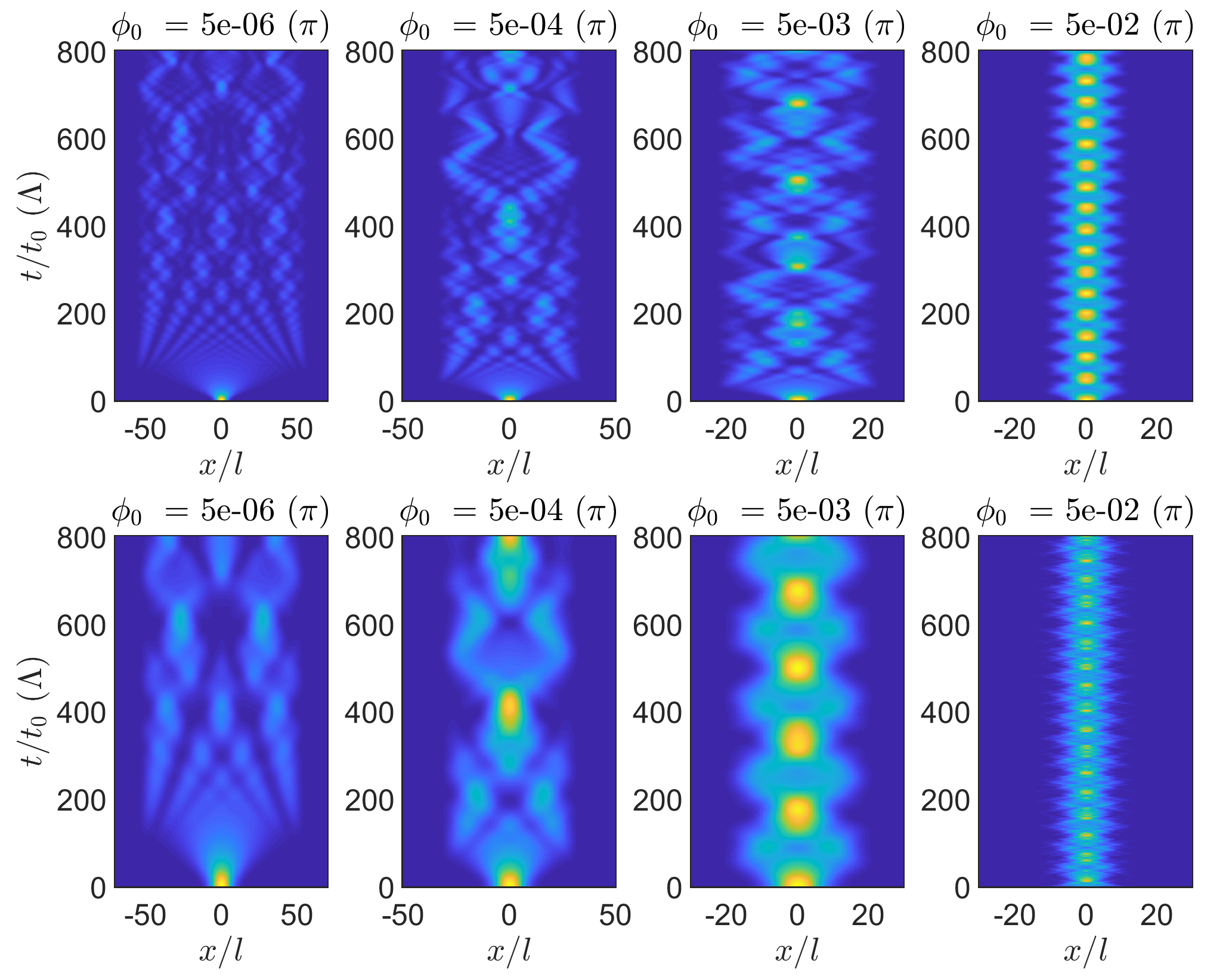} 
\caption{\label{fig:odd_delay} {Intensity distribution for a sine hyperbolic delay $\phi(x)$ given by Eq.~\eqref{eq:delay_odd}. The incident wave is Gaussian shaped, with $w_G/l=5$ and $w_G/l=10$ on the top and bottom row, respectively. Each column corresponds to a different depth of the potential as dictated by the parameter $\phi_0$ (reported in units of $\pi$ on the top of each panel): the confinement is thus enhanced from left to right. The longitudinal period is fixed to~$\Lambda$, whereas the spatial profile $U(x)$ is a flat-top of normalized width 2000 and amplitude 0.5. Here $\varphi_\mathrm{in}=\pi/2$.}}
\end{figure*}

Figure~\ref{fig:odd_potential_vs_max_phase} shows the numerical results for a fixed width of the potential $w_\mathrm{well}/l=10$ and increasing values (from top to bottom) of the maximum phase delay $\phi_\mathrm{max}$. For a flat distribution of delay, the effective potential is bell-shaped but repelling, yielding a symmetric splitting of the input wave into two branches propagating in opposite transverse directions [Fig.~\ref{fig:odd_potential_vs_max_phase}(a)]. For small enough $\phi_\mathrm{max}$, the value of the effective potential remains positive around the origin $x=0$, see Fig.~\ref{fig:odd_potential_vs_max_phase}(b). Repulsion is even increased with respect to the previous case of vanishing phase delay [plotted in Fig.~\ref{fig:odd_potential_vs_max_phase}(a)], in turn yielding a faster and a parity-broken deflection of the wave. For further increase in $\phi_\mathrm{max}$ the contribution due to the delay in $V_\mathrm{eff}$ becomes dominant, with the wave now, on average, being trapped around the origin, see Fig.~\ref{fig:odd_potential_vs_max_phase}(c,d). The odd potential $U(x)$ is manifested as a continuous oscillation on the $x$-direction on the fast scale. This fast wiggling motion is periodic for intermediate values of the delay [Fig.~\ref{fig:odd_potential_vs_max_phase}(c)], whereas it becomes aperiodic for large enough delays $\phi$ [Fig.~\ref{fig:odd_potential_vs_max_phase}(d)].\\
Next, we investigate how the wave evolution depends on the width $w_\mathrm{well}$ of the potential shape. Results for a large value of $\phi_\mathrm{max}$ (specifically $\phi_\mathrm{max}=2\pi$) are plotted in Fig.~\ref{fig:odd_potential_vs_wwell}, where the amplitude of the potential $U(x)$ is kept constant and only the normalized width $w_\mathrm{well}/l$ is varied. In this case the contribution proportional to $\left(\partial \phi/\partial x\right)^2$ in the definition of the effective potential $V_\mathrm{eff}$ [see Eq.~\eqref{eq:Veff}] is dominant (at lower $\phi_\mathrm{max}$ an interplay is occurring, see Fig.~\ref{fig:odd_potential_vs_max_phase}). Nonetheless, the shape of the potential $U(x)$ is still affecting $V_\mathrm{eff}$ by directly modulating $\left(\partial \phi/\partial x\right)^2$ via a factor $U^2(x)$, in this case the latter being shaped as a bell-shaped function. Thus, for narrow potentials (small $w_\mathrm{well}$) the confinement due to the transversely-dependent delay improves (compare the variations in amplitude of $V_\mathrm{eff}$ in Fig.~\ref{fig:odd_potential_vs_wwell} from top to bottom panels). While the wave is trapped in the central lobe of the effective potential, the odd symmetry of the underlying instantaneous potential $U(x)$ is manifesting as a local wiggling inside the central lobe itself [Fig.~\ref{fig:odd_potential_vs_wwell}(a,b)]. As the potential widens, both the transverse confinement and the fast transverse motion decreases, yielding a long-period (with respect to the period $T$) breathing, including a slight left-right asymmetry in the wave profile. For even wider potentials, the trapping effect keeps decreasing, resulting first in wider beams and longer breathing period [Fig.~\ref{fig:odd_potential_vs_wwell}(c)], and finally in the lack of trapping around the origin [Fig.~\ref{fig:odd_potential_vs_wwell}(d)].

\section{Odd-symmetric delay for an even instantaneous potential}
\label{sec:odd_delay}
Lastly, we analyze the case when the delay $\phi(x)$ is odd symmetric. We also consider a flat-top profile much larger than the support of the delay $\phi(x)$, so that Eq.~\eqref{eq:Veff_only_delay} can be safely applied. Despite that, the effective potential $V_\mathrm{eff}(x)$ is even symmetric. For shapes of the delay presenting larger absolute values of the derivatives $|d\phi/dx|$ in the center than in the edges (for example a hyperbolic tangent), the effective potential becomes repelling. Much more interesting is the opposite case. As a matter of fact, when the absolute value of the derivative $|d\phi/dx|$ in the center is smaller than in the edges, the effective potential features a global minimum around $x=0$, that is, localized quasi-modes can exist. To investigate the wave trapping in this case, we take for the delay a hyperbolic sine function
\begin{equation} \label{eq:delay_odd}
  \phi(x)=\phi_0 \sinh\left(\frac{x}{w_\phi}\right).
\end{equation}
From Eq.~\eqref{eq:delay_odd}, %  $\phi(x=w_\phi)=\left[\left(e-e^{-1}\right)/2\right]\phi_0 \approx 1.17\ \phi_0$.
$\phi(x=w_\phi)=\sinh(1)\ \phi_0 \approx 1.17\ \phi_0$.
 Applying Eq.~\eqref{eq:Veff_only_delay}, the effective potential reads
\begin{equation} \label{eq:Veff_odd_delay}
  V_\mathrm{eff} \approx \frac{T^2 V_\mathrm{cen}^2 \phi_0^2}{16 \pi^2 m w_\phi^2}\cosh^2\left(\frac{x}{w_\phi}\right),
\end{equation}
Figure~\ref{fig:odd_delay} shows the wave propagation for two different Gaussian inputs ($w_G/l=5$ and $w_G/l=10$ on the top and bottom row respectively) when ${w_\phi}/{l}=5$ and for various values of $\phi_0$ (labelled at the top of each panel). The shape of the potential is very similar to a parabolic one, with the important feature to be infinitely extended (and therefore infinitely deep) along $x$. Therefore, in the range of validity of the model discussed here, the structure confines on the transverse plane the impinging wave, regardless of its profile. This behavior is similar to a metallic waveguide in the optical case. Such a behavior is confirmed by the numerical simulations, plotted in Fig.~\ref{fig:odd_delay}. For different widths of the Gaussian input, the wave keeps being localized around $x=0$, in a region determined by the amplitude ${\phi_0^2}/{w_\phi^2}$. Noteworthy, complicated interference patterns emerge in time due to the simultaneous excitation of several spatial quasi-modes, in accordance with the potential \eqref{eq:Veff_odd_delay}.\\
\begin{figure}
\includegraphics[width=0.45\textwidth]{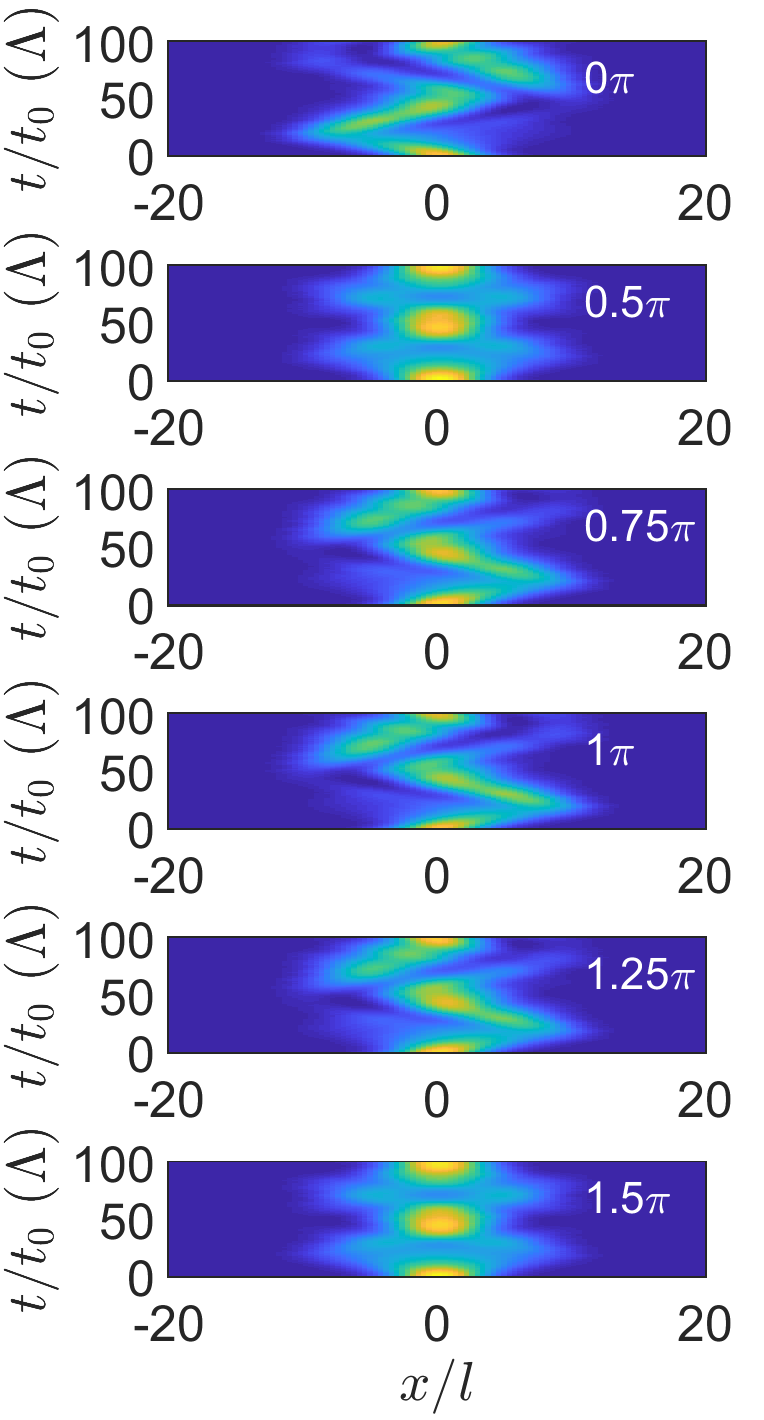} 
\caption{\label{fig:odd_delay_shift} {Intensity distribution for a sine hyperbolic delay $\phi(x)$ given by Eq.~\eqref{eq:delay_odd}. The incident wave is Gaussian shaped featuring $w_G/l=5$. Each row corresponds to a different temporal shift $\varphi_\mathrm{in}=0,\ 0.5\pi,\ 0.75\pi,\ \pi,\ 1.25\pi,\ 1.5\pi$, from top to bottom. The longitudinal period is $\Lambda$ and $\phi_0=5\times 10^{-2}\pi$, whereas the spatial profile $U(x)$ is a flat-top of width 2000$l$ and amplitude 0.5.}}
\end{figure}
As pointed out by Eq.~\eqref{eq:quasimode}, the wavefronts of the quasimodes are non-planar and varying continuously in propagation: as a matter of fact, a strong dependence on the initial phase $\varphi_\mathrm{in}$ of the longitudinal variation of the potential is expected \cite{Alberucci2013}. The numerical simulations for different $\varphi_\mathrm{in}$ are shown in Fig.~\ref{fig:odd_delay_shift}. The initial condition for the wave is kept fixed, whereas the function $f(t)$ is shifted in time. By specializing Eq.~\eqref{eq:quasimode} to our case, the quasi-mode in $t=0$ reads
\begin{equation}  \label{eq:quasimode_odd_delay}
  \psi(x,t)\approx g(x) e^{-i\frac{V_\mathrm{cen}T}{2\pi\hbar}\sin\left[\varphi_\mathrm{in} - \phi(x)\right]}. %e^{-i\frac{E_0 t}{\hbar}}.
\end{equation}
The parity of the mode then depends on the potential shift $\varphi_\mathrm{in}$: when $\varphi_\mathrm{in}=\pi/2+k\pi\ \left(k\in \mathcal{Z} \right)$, the quasi-mode is even, albeit with a non-flat phase front; for other values of $\varphi_\mathrm{in}$, there is an odd component in the phase profile, yielding a qausi-periodic  transverse oscillation of the wave in propagation. The predicted dynamics is confirmed by the results shown in Fig.~\ref{fig:odd_delay_shift}.
The wave dynamics in time strongly changes with the initial phase: intensity profiles range from transverse oscillations (odd symmetric) to straight trajectories (even symmetric) for $\varphi_\mathrm{in}=0.5\pi$ and $1.5\pi$. Regardless of the temporal shift, the wave being confined always in the same region of space determined by the effective potential $V_\mathrm{eff}$.

\section{Experimental Verification}
\label{sec:fiber_loop}

\begin{figure}
\includegraphics[width=0.49\textwidth]{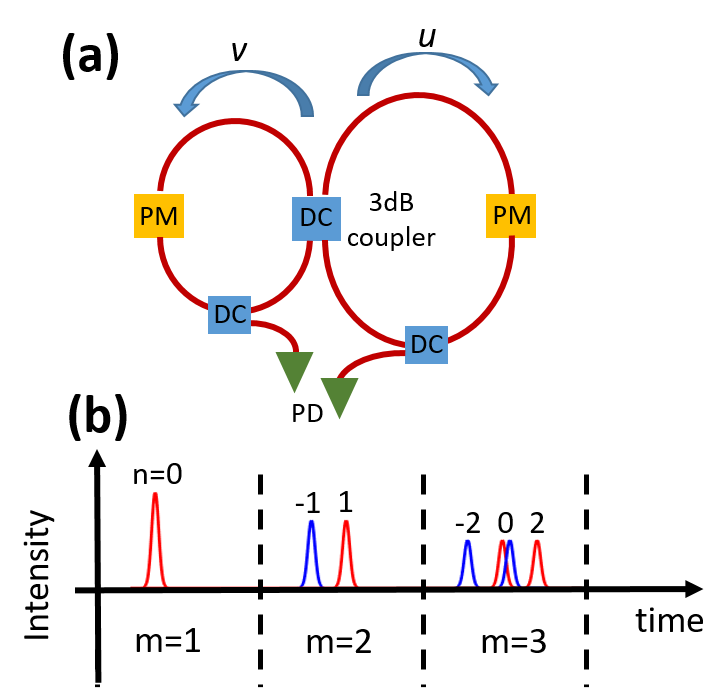} 
\caption{\label{fig:setup}{ (a) Simplified schematic of the experimental set-up: Two fiber loops of different lengths are interlaced via a 3~dB directional coupler (DC). A phase modulator (PM) is placed in each loop. %; in our experiments the two modulations are identical at each instant. 
Two additional DCs extract a portion of the circling energy for real time measurement of $u$ and $v$ via two photo-detectors (PDs). Amplifiers (not shown) are inserted to compensate for the losses. (b) The modulation scheme: a single initial pulse is injected into the the loop $u$ (represented by the red curves) at $m=1$; after a round trip, at $m=2$ the 3dB coupler splits the pulse into two halves, one per loop (pulses in $v$ are plotted in blue). The duplication repeats at each additional round trip (see $m=3$ in the figure). Now, at the same time $n=0$ (in the plot the two pulses are slightly shifted for the sake of clarity) two pulses of same amplitude but different phases are propagating, one per loop. }}
\end{figure}

\begin{figure*}
\includegraphics[width=0.99\textwidth]{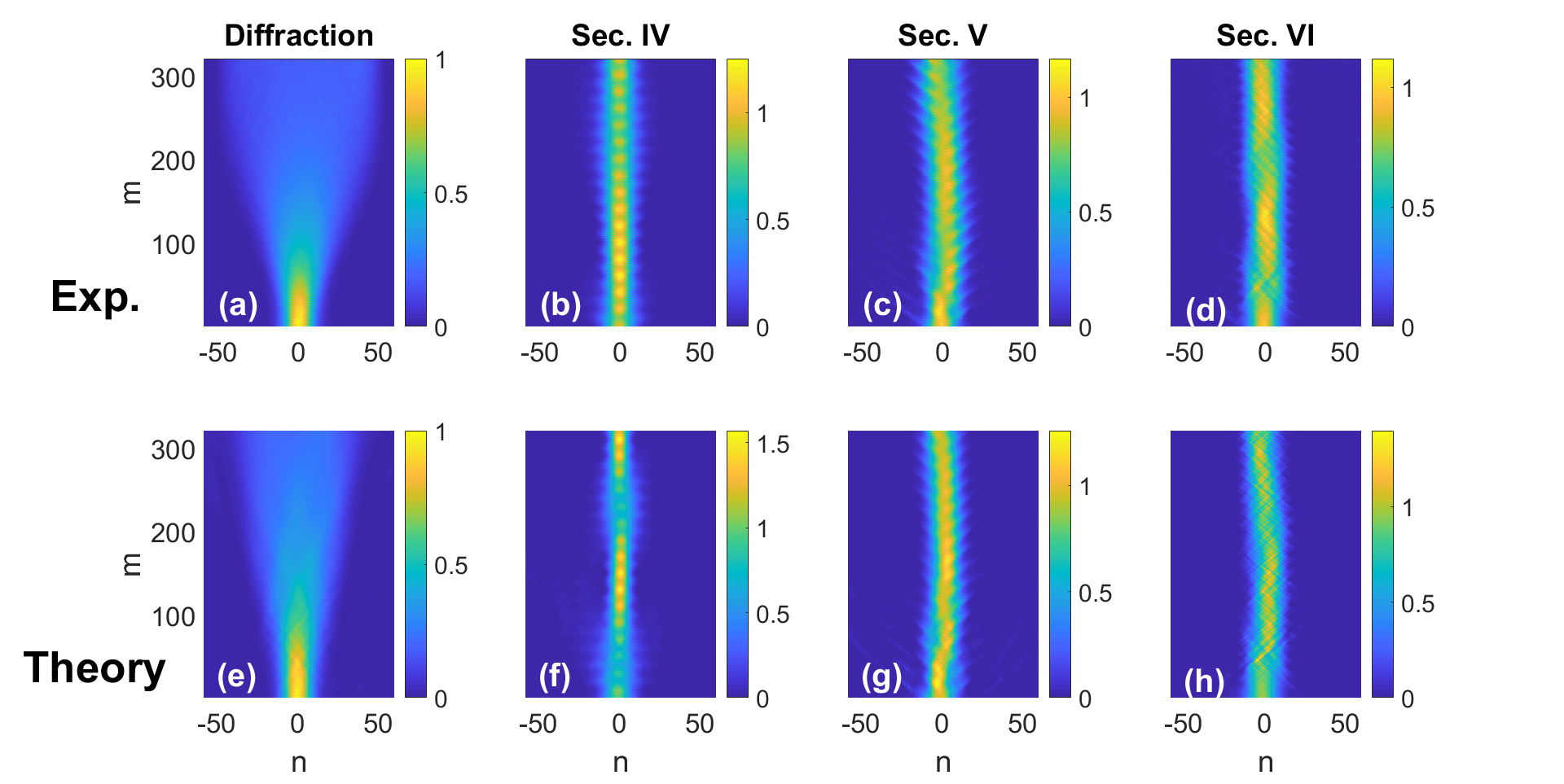} 
\caption{\label{fig:fiber_loop}{ Experimental verification in fiber loop for a modulation period of 20. Experimental (top row) and simulated (bottom row) intensity distribution of the diagonal component $\propto \bm \varphi_+$ on the plane $nm$. (a,e) Propagation without phase modulation. (b,f) Confinement for 
 $U(n)=-\pi \exp{\left[-\left(n/w_G\right)^8\right]}$ with $w_G=100$, $\phi=-0.5\pi \exp{\left[-n^2/w_\phi^2 \right]}$ with $w_\phi=30$. (c,g) Confinement for $U(n)=0.2\pi \tanh{\left[\left(n/w_G\right)\right]}$ with $w_G=30$, $\phi=4\pi \exp{\left[-n^2/w_\phi^2 \right]}$ with $w_\phi=30$. (d,h): Confinement for $U(n)=0.2\pi \exp{\left[-\left(n/w_G\right)^4\right]}$ with $w_G=50$, $\phi=0.025\pi \sinh{\left(n/w_\phi \right)}$ with $w_\phi=5$. The same diagonally polarized input $\bm \varphi_+$ is taken both for the experiments and the simulated profiles. }}
\end{figure*}
To experimentally verify the theoretical results presented in the above sections, we refer to discrete optical lattices based upon fiber loops \cite{Regensburger:2011}. The basic idea of the set-up is to discriminate different pulses according to their positions in time \cite{Schreiber:2010}. Beyond being used for the demonstration of the Kapitza effect \cite{Muniz:2019_1}, fiber loops have been employed for investigating several basic physical phenomena, such as $PT$ symmetry breaking \cite{Miri:2012,Regensburger:2012}, diametric drive \cite{Wimmer:2013}, Berry curvature in periodic systems \cite{Wimmer:2017}, topological effects  \cite{Bisianov:2019,Weidemann:2020,Weidemann:2022}, superfluidity \cite{Wimmer:2021}, machine learning \cite{Pankov:2022}, and so on. The experimental configuration and the basic principle are shown in Fig.~\ref{fig:setup}.  The set-up is basically composed of two fibers of slightly different length configured in closed loops, interacting with each other through a standard directional coupler (DC). Without any external modulation, at each round trip a single pulse splits into two copies, the relative weight being fixed by the DC. Here we apply a 3~dB DC, i.e., the same amount of power on each replica. The different length of each loop translates into a different arrival time for each of the copies at the following passage through the DC. In essence, each round plays the role of a discrete evolution coordinate (time $t$ above), whereas the temporal separation between each delayed replica in a single round trip plays the role of the transverse coordinate $x$. Accordingly, the lengths of the two fibers are chosen to ensure that the pulses from different round trips (i.e., different synthetic time $t$) are not overlapping inside our observation window (approximately of 300 loops, see below). By adding phase modulators along the fibers, an external electromagnetic field is effectively acting on the propagation of the discrete train of pulses. \\ %In this paper we will refer to the case where the same phase modulation is applied in each loop, realized by inserting a phase modulator of identical output in each half of the system.
Calling $u$ and $v$ the field amplitude in the long and short loop respectively, the evolution can be written as \cite{Wimmer:2013,Muniz:2019_1}
\begin{align}
   \begin{split}
   & u_{n}^{m+1}= \frac{1}{\sqrt{2}} (u_{n-1}^{m} + i v_{n-1}^{m})e^{i\Phi_u(m,n)},  \\
   &  v_{n}^{m+1}= \frac{1}{\sqrt{2}} (v_{n+1}^{m} + i u_{n+1}^{m})e^{i\Phi_v(m,n)}.
   \label{eq:uv}
   \end{split}
\end{align}
In Eq.~\eqref{eq:uv} the indices $m$ and $n$ represent the longitudinal and transverse coordinates, respectively. The phase modulators inserted in the loop $u(v)$ impart a phase delay $\Phi_{u(v)}$. To make Eq.~\eqref{eq:uv} similar to a \schro equation, we set $\Phi_u=\Phi_v=\Phi(n,m)$. In the continuous limit, we define the two-component variable $\bm{\varphi} = (u;\ v)$. We further introduce $\bm \varphi_{\pm}= \left(u; \pm v \right)/\sqrt{2}$ as the pseudo diagonal and anti-diagonal polarizations, respectively. After setting $\bm \varphi =\psi(n,m) \bm \varphi_{\pm} $, Eq.~\eqref{eq:uv} in the limit of small transverse momenta can be recast as a pair of independent \schro equations \cite{Muniz:2019_1,Wimmer:2021}
\begin{equation}
i\frac{\partial \psi}{\partial m}
 = \pm\left( \frac{\pi}{4} -  \frac{1}{2} \frac{\partial^2}{\partial n^2}\ \right) \psi +  \Phi(n,m) \psi , \label{eq:Continuous}   
\end{equation}
In Eq.~\eqref{eq:Continuous} the $\pm$ sign corresponds to a different sign of the mass due to the presence of bands of different curvatures in the original discrete system \cite{Wimmer:2013,Lechevalier:2021}.  Finally, from Eq.~\eqref{eq:Continuous} it is evident that  the imparted phase $\Phi$ plays the role of the scalar potential $V(x,t)$ in Eq.~\eqref{eq:SE}. \\
Experimentally speaking, a Gaussian-shaped train of coherent pulses with the diagonal pseudo-polarization $\bm \varphi_+$ in the form $\psi=\exp{\left(-n^2/w_\mathrm{in}^2\right)}$ [see Eq.~\eqref{eq:gaussian_input}] is prepared from a single initial pulse by a proper choice of the phase modulation \cite{Wimmer:2013}; in our experiments it is $w_\mathrm{in}=12.5\pm 0.5$. Whereas ideally the input should be fully polarized along $\bm \varphi_+$, in our set-up there is a spurious component $\bm \varphi_-$ carrying about $10\%$ of the overall injected energy. Experimental results are plotted in the top row of Fig.~\ref{fig:fiber_loop}. The corresponding theoretical results, shown in the bottom row of Fig.~\ref{fig:fiber_loop}, are calculated using Eq.~\eqref{eq:uv} with the input used in the experiment, thus accounting for the presence of noise generated during the initial selection of the Gaussian train. First, in Fig.~\ref{fig:fiber_loop}(a) we verify the dispersive spreading of the beam in the absence of any phase modulation. Direct comparison with the theoretical predictions shows a good agreement, with the wave widening being very similar. From the distribution of the other component (not shown), we can evince a slight asymmetry between positive and negative $n$ due to a small imbalance in the two fiber loops employed in the experiments, thus explaining the observed small differences. We then demonstrate the wave confinement by comparing the diffracting case plotted in Fig.~\ref{fig:fiber_loop}(a) with the pulse distribution acquired when a retarded potential is applied via the phase modulators. For all the  different symmetries presented in the previous sections (the title of each column provides the corresponding section of the paper where the given type of potential and delay is investigated), a clear transverse trapping of the wave is observed. Multiple repetition of the experiments demonstrate that the confinement is robust to the input noise. The theoretical results agree well with the experiments. We stress that, owing to the asymmetry in two fiber loops discussed above, in the experiments a small but net increase of power is occurring versus $m$. To compensate for this drifting of the set-up, the experimental results plotted in Fig.~\ref{fig:fiber_loop}(a-d) are normalized with respect to the power transported at each cross section $m=\text{constant}$. \\ % Larger discrepancies are observed in Fig.~\ref{fig:fiber_loop}(b,f).
\begin{figure}
\includegraphics[width=0.5\textwidth]{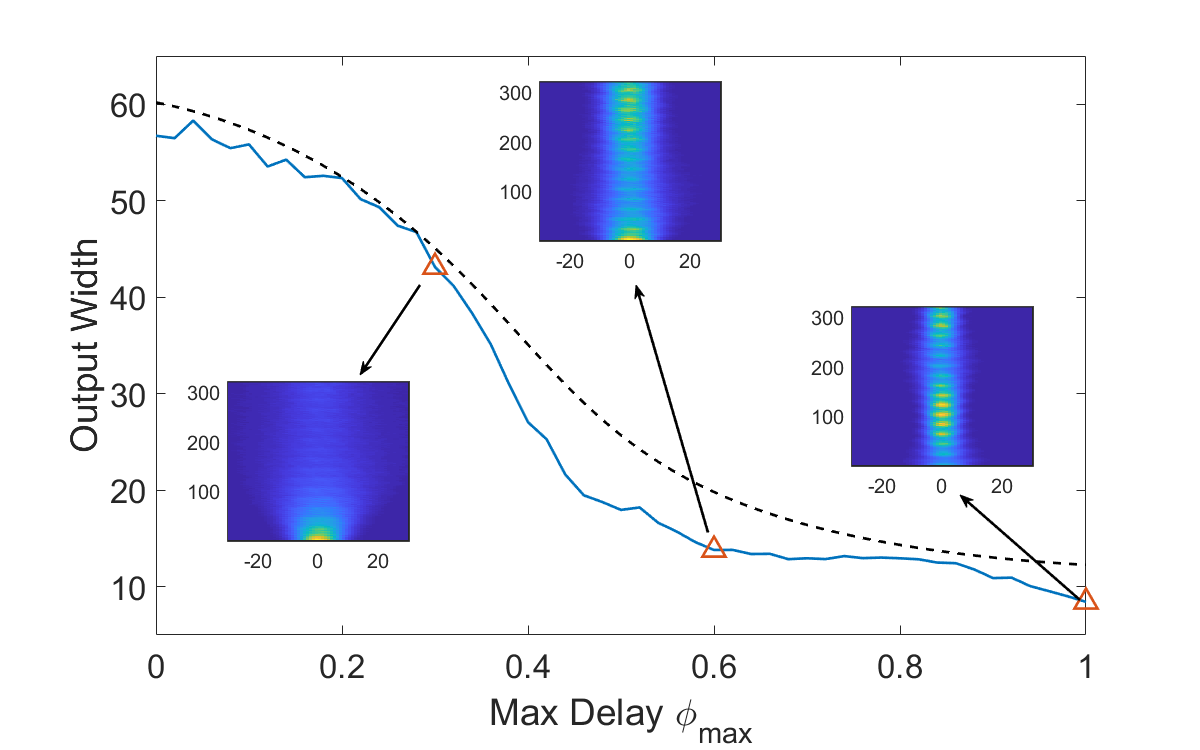} 
\caption{\label{fig:confinement_vs_delay_exp}{ Beam width at $m=380$ versus the maximum of the delay in the case of even potential and even delay, the latter of Gaussian shape. Solid and dashed lines are the experimental and theoretical results, respectively. The insets show the distribution of the pulses on the plane $nm$ for the cases indicated with a triangle on the experimental curve. }}
\end{figure}
For the sake of completeness, we investigate the evolution of the wave trapping as the delay is increased for the case of an even potential and even delay distribution discussed in Sec.~\ref{sec:pot_even_delay_even}. We thus chose $\phi(n)=\phi_\mathrm{max}\exp{\left(-n^2/w_\phi^2\right)}$ as the delay profile applied on the super-Gaussian potential and selected a fixed Gaussian input of width $13.5\pm 0.5$ at the entrance, the uncertainty on the input width being ascribed to the experimental noise. We investigate the light propagation as $\phi_\mathrm{max}$ increases: Figure~\ref{fig:confinement_vs_delay_exp} compares the experimental (solid blue line) and the calculated width (black dashed line) of the beam at $m=380$ for a Gaussian input of width 13.5. The confinement of the wave due to the gauge field is clear, with a gradual transition from diffraction to localization. The insets display the field distribution for some particular cases, showing how the effective guide becomes multimodal for large $\phi_\mathrm{max}$, as proved by the appearance of a breathing dynamics in the intensity distribution.

%\cite{Chalabi:2020}

\section{Discussion of the results and their applicability}
\label{sec:discussion}
Being based on the \schro equation, our mathematical results can be applied (or generalized by using the proper Lagrangian for the light-matter interaction) to several physical settings. One potential application is the Kapitza-Dirac effect (KDE), where a matter wave (in the original proposal composed by charged particles such as electrons) in free propagation is scattered by a stationary optical wave \cite{Batelaan:2000}. Whereas such an effect can be both explained quantum mechanically and classically \cite{Chan:1979,Batelaan:2000}, in the regime of large intensities the KDE can be effectively modelled using the ponderomotive force, the latter being proportional to the gradient of the local average intensity \cite{Bucksbaum:1988}. Our model can inspire new scattering potentials by inserting for example an inhomogeneous phase mask in the path of the standing wave, thus generating a temporal shift in the instantaneous intensity distribution. In such a way, new modulation schemes for the impinging electron beam could be designed \cite{Dahan:2021}.  \\
In full analogy with the KDE, our results can be applied to optical traps, i.e., properly tailored laser beams capable to trap neutral particles with certain spatial patterns \cite{Otte:2020}. %In that case, the electromagnetic field provides the periodic potential $V(x,t)$. 
 Our findings can pave the way to a new set of optical traps based upon the modulation of the optical phase distribution. As a matter of fact, the seminal work of Ashkin and the previous fundamental works on light-matter interaction showed that an electromagnetic force %\textendash called ponderomotive force \textendash 
induces a force proportional to the intensity gradient, in agreement with the Kapitza potential \cite{Gordon:1973,Ashkin:1978}. Another field where the Kapitza theory can be applied is nonlinear optics, specifically to explain a non-perturbative effect such as high-harmomic generation (HHG). In HHG, a strong ultrashort optical pulse extracts electrons from a material (via tunnelling or multiphoton absorption); the excited electrons then behave like a quasi-free particle whose motion is mainly dictated by the incident electromagnetic field \cite{Lewenstein:1994}. In this regime the Kapitza approach can find application, as recently addressed in Ref.~\cite{Zheltikov:2019}. As a matter of fact, our results can be generalized (i.e., using the complete Lagrangian accounting for the light-matter interaction, see e.g. Ref.~\cite{Watzel:2020} for an approach including complex light states) in order to understand how structured light beams can affect HHG, with impact in related fields such as the generation of attosecond pulses \cite{Kozak:2018}.\\
The role between matter and light can also actually been exchanged, that is, light being considered like the field subject to a potential dependent on the matter configuration. As already discussed in the introduction, it is well known that light propagation in the harmonic regime is ruled by a \schro equation with the propagation distance playing the role of an effective time \cite{Longhi:2009}, an analogue massively exploited in the last decades to demonstrate general physical effects using relatively simple optical set-ups \cite{Schwartz:2007,Rechtsman:2013,Cardano:2015,Marcucci:2019,Cerjan:2020}. Following this lead, a time-periodic potential corresponds to a longitudinally periodic distribution of the refractive index. These structures can be realized in several technological platforms, ranging from integrated optics \cite{Li:1992,Yu:2009,Bock:2010} to direct laser writing in transparent materials \cite{Fernandes:2011,Phillips:2015}. In particular, our results can be potentially applied to the realization of a new type of 3D resonators, simultaneously ensuring longitudinal trapping (through the Bragg reflections stemming from the longitudinal periodicity) and transverse trapping due to the Kapitza effect. Such kind of structures are at reach for example using volume Bragg gratings inscribed via ultrashort optical pulses \cite{Talbot:2020}. \\
Our results are also strictly connected to the realization of photonic devices based upon the gauge field, in particular a novel type of waveguides. The idea to use synthetic gauge fields for light confinement has been initially introduced by Lin and Fan in Ref.~\cite{Lin2014a}, where a lattice made of resonators had been proposed to realize the idea experimentally. The geometry proposed in this article presents an easier implementation of this idea by using a continuous system. Noteworthy, gauge-induced trapping has been recently demonstrated by Lumer and collaborators using arrays of femtosecond-written waveguides in fused silica using two different configurations \cite{Lumer:2019}. In particular, they have demonstrated that in this discrete system light can be spatially localized in the presence of a longitudinally-periodic coupling between waveguides, the latter being shifted on the transverse plane of the array. Our model gives a more intuitive explanation for such effects with respect to the Bloch band approach used by Lumer in waveguide arrays \cite{Longhi2011}; finally, our approach provides a generalization of the basic concept to the continuous case, the latter being of more interest and potential impact in real world applications.

\section{Conclusions}
\label{sec:conclusions}
We investigated the propagation of dispersive waves in the presence of a periodic potential encompassing a null average, the potential being subject to a space-dependent delay. We framed the action of an inhomogeneous delay in the context of gauge transformations. The influence of the local delay on the wave evolution has been modelled by introducing an effective time-independent potential: the effective time-independent potential consists of a Kapitza potential featuring an additional contribution due to the delay. With these tools in hand, we first verified the existence of wave confinement by a proper tailoring of the delay. We then discussed the strong impact played by the parity symmetry of the instantaneous potential in determining the wave propagation in time, emphasizing the complex interplay between the fast micromotion and the behavior of the averaged field on the slow scale. We finally verified our predictions in fiber optical loops, where a gauge-dependent confinement is demonstrated for each case discussed above.\\
Due to the generality of our model, our results can find application in several fields, including cold atoms, optical traps, high harmonic generation, and photonics, among the others. Future generalizations include the possibility to induce an effective magnetic field when two transverse dimensions are accounted for, and the exploration of more sophisticated spatio-temporal modulations of the potential and the related relationship with apparent forces \cite{Longhi2007,Bliokh2015,Bliokh:2019}.

\appendix

\section{Normalization}
\label{sec:normalization}
Setting $\xi=x/l$ Eq.~\eqref{eq:SE} reads
\begin{equation} \label{eq:SE_2}
 i\frac{\hbar l^2}{D} \frac{\partial \psi}{\partial t}= -\frac{\partial^2 \psi}{\partial \xi^2} + \frac{V(\xi,t)l^2}{D} \psi. \nonumber
\end{equation}
Defining the normalized quantities $\zeta=tD/\left(\hbar l^2 \right)$ and $W(\xi,\zeta)=V(\xi,\zeta)l^2/D$, Eq.~\eqref{eq:SE_2} provides
\begin{equation}
   i\frac{\partial \psi}{\partial \zeta}= -\frac{\partial^2 \psi}{\partial \xi^2} + {W(\xi,\zeta)}{} \psi \label{eq:schro_normalized}
\end{equation}
Equation~\eqref{eq:schro_normalized} is adimensional. By the definition of the normalized potential $W(\xi,\zeta)$, in the case of longitudinally-invariant potentials the profiles of the bound modes are dictated by the ratio between the potential and the diffraction coefficient. The normalized dispersion coefficient $D/l^2$ provides a scaling coefficient with respect to the phase in propagation: the higher the dispersion, the larger the propagation constants are.  The effective Kapitza potential then reads
\begin{equation} \label{eq:Veff_normalized}
 W_\mathrm{eff}(\xi,\zeta) = %\left(\frac{\hbar l^2}{D}\right)^2 \frac{D}{\hbar^2} \left(\frac{D}{l^2}\right)^2 
\frac{D T_\zeta^2}{l^2} \sum_{n\neq 0} \frac{1}{4\pi^2 n^2} 
\left| \frac{\partial V_n}{\partial \xi}\right|^2, 
\end{equation}
%\begin{equation}
% W_\mathrm{eff}(\xi,\zeta) = \left(\frac{\hbar l^2}{D}\right)^2 \frac{D}{\hbar^2} \left(\frac{DT_\zeta}{l^2}\right)^2 \frac{1}{l^2} \sum_{n\neq 0} \frac{1}{4\pi^2 n^2} 
%\left| \frac{\partial V_n}{\partial \xi}\right|^2 
%\end{equation}
where we set $T_\zeta=T D/\left(\hbar l^2 \right)$. Thus, unlike the case of time-independent potentials, the Kapitza effective potential depends on the normalized dispersion $D/l^2$.
For a given potential $V(x/l)$, the Kapitza effect is greater for larger $D$: the mass of the particle is smaller and it is easier (with respect to heavier particles) to accumulate kinetic energy. Analogously, the effective potential is stronger for smaller $l$, the latter corresponding to stronger forces acting on the waves for a fixed potential drop. Finally, we notice that the dependence from $D$ and $l$ disappears once $W_n$ is used in Eq.~\eqref{eq:Veff_normalized}, in accordance with Eq.~\eqref{eq:schro_normalized}.

\section{Higher order terms in the Kapitza model}
\label{sec:model_accuracy}
The formula for the effective potential holds valid only up to the order $O(T^2)$, as explained in detail in Ref.~\cite{Alberucci:2020}. Here we briefly revise the basic method to find the higher order terms. 
The complete expansion of the field reads \cite{Alberucci:2020}
\begin{equation}
    \psi(x)=h(x)\exp{\left[-\frac{i}{\hbar}\sum_{n=0}^\infty  e_n(x) \int_{t_\mathrm{in}}^{t}{e^{-\frac{2\pi i n t}{T}} dt} \right]}, \label{eq:integral_series}
\end{equation}
where $t_\mathrm{in}$ is the initial instant and $e_n(x)$ are point-dependent pseudo-energies. To obtain the function $g(x)$ used in the main text, we need to eliminate the dependence from the initial time $t_\mathrm{in}$ using the additional transformation % is the eigenfunction of the \schro equation with a time-independent potential given by Eq.~\eqref{eq:Veff_delay}. The connection with the function $g(x)$ is given by the gauge transformation
\begin{equation}
    h(x) = g(x) \exp{\left[-\frac{i}{\hbar}\sum_{n=0}^\infty \left. e_n(x)\int{\exp{\left(-\frac{2\pi i n t}{T}\right)} dt} \right|_{t_\mathrm{in}}\right]}.
\end{equation}
We then make the expansion $e_n(x)= e_n^{(0)}(x) + e_n^{(1)}(x)T + O(T^2)$. The terms $e_n^{(k)}$  can be found by direct substitution into the \schro equation;
at the leading order, it is $e_n^{(0)}(x) = U(x)e^{-2\pi i n \frac{\tau(x)}{T}}$.  The main difference with respect to the standard Kapitza case (i.e., an $x$-independent delay) is that the functions $e_n(x)$ are now featuring a non-flat wavefront along $x$ at any level of approximation. In the case of a sinusoidal modulation, we find Eq.~\eqref{eq:quasimode}. 
%The intrinsic complex nature of $e_n^{(0)}(x)$ implies a different behavior of the higher order terms when a point-wise delay is added. 
The complete formula for the Kapitza potential reads \cite{Alberucci:2020}
\begin{equation}
  V_\mathrm{eff}(x)=\frac{T^2}{8\pi^2 m}\sum_n \frac{1}{n^2}\left( \frac{\partial e_n}{\partial x} \frac{\partial e_{-n}}{\partial x} \right). \label{eq:complete_Kapitza}
\end{equation}
Once the perturbative series for $e_n(x)$ is inserted into Eq.~\eqref{eq:complete_Kapitza}, the effective potential can be then expressed in a power series of the period $T$, thus generalizing Eq.~\eqref{eq:Veff} to the case of long periods with respect to the Rayleigh distance. The behavior of the quasi-mode in one period can eventually be calculated by direct substitution of $e_n(x)$ into the initial ansatz given by Eq.~\eqref{eq:integral_series}.
%The next step is to introduce the power expansion of $e_n(x)$ into Eq.~\eqref{eq:complete_Kapitza}. In the standard case, the effective potential $V_\mathrm{eff}$ remain real up to the $e_n^{(2)}(x)$ term. In the current case, the non-flat phase profile for $e_n^{(0)}(x)$ implies a complex effective potential already when terms $e_n^{(1)}(x)$ are taken into account in Eq.~\eqref{eq:complete_Kapitza}. Hence, the quasi-mode is now the solution of a complex-valued eigenvalue problem. Numerical simulations confirm that the direct application of the method presented in Ref.~\cite{Alberucci:2020} does not lead to an improvement in the confinement of the wave.

\begin{figure*}
\includegraphics[width=0.95\textwidth]{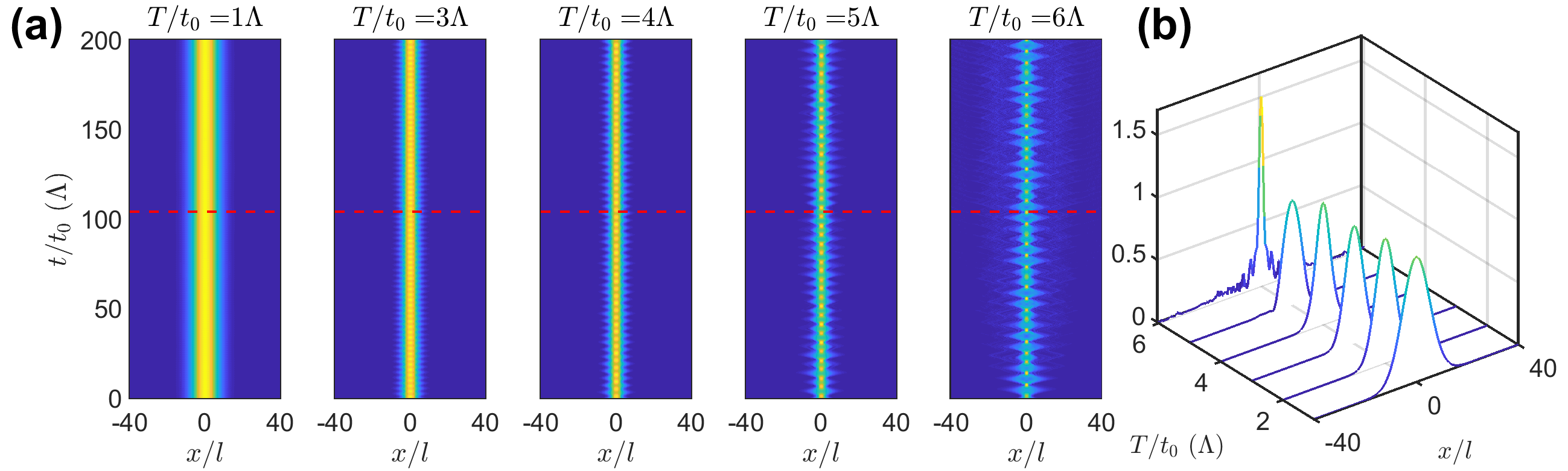}
\caption{\label{fig:numerics_parabolic} {Propagation of the quasi-mode in the case of full confinement for a potential given by Eq.~\eqref{eq:potential_full_confinement} with $x_0/l=40$, $N=16$, and $al^2=9.8\times 10^{-3}$. The potential in propagation is sinusoidal with $U(x)=0.5e^{-x^4/w_\mathrm{pot}^4}$ and $w_\mathrm{pot}/l=2800$. In (a) $|\psi|^2$ on the propagation plane is plotted for increasing periods $T/t_0$ from left to right, when in the input condition the phase front curvature is accounted for. In (b) the modulus of the wave $|\psi|$ calculated in $t/t_0=104\Lambda$ [slice corresponding to the red dashed line in (a)] is plotted versus $x/l$ for periods $T/t_0$ ranging from $\Lambda$ to 6$\Lambda$. Here $\varphi_\mathrm{in}=0$.}}
\end{figure*}

\section{Designing a V-shaped effective potential}
A full confinement of the quasi-mode can be achieved also by using a V-shape distribution for the local delay. We use the following ansatz for $\phi(x)$ \cite{Jisha2017}
\begin{equation} \label{eq:potential_full_confinement}
 \phi= \frac{1}{2} a x^2 M_N\left(\frac{x}{x_0}\right) + \left( a x_0 |x| -\frac{1}{2}ax_0^2 \right) \left[ 1- M_N\left(\frac{x}{x_0}\right)\right], 
\end{equation}
where $M_N(x)=e^{-x^N/x_0^N}$ is the super-Gaussian function of order $N$. The effective potential $V_\mathrm{eff}$ corresponding to Eq.~\eqref{eq:potential_full_confinement} in the limit $N\rightarrow\infty$ is parabolic for $|x|<x_0$ and constant elsewhere. Thus, fully localized modes (with reference to the Kapitza model) are supported given that $\lim_{|x\rightarrow\infty|}V_\mathrm{eff}>V_\mathrm{eff}(x=0)$. Numerical results are shown in Fig.~\ref{fig:numerics_parabolic}, showing a good degree of spatial localization up to $T/t_0=6\Lambda$, thus showing better performances than obtained for Gaussian delay (compare with Fig.~\ref{fig:numerics_gaussian}). For longer periods confinement disappears due to the failure of the Kapitza model due to the significant contribution of the higher harmonics.

\section{The optical case}
\label{sec:optical_case}
In the optical case Eq.~\eqref{eq:SE} corresponds to the paraxial Helmholtz equation after the substitution $t\rightarrow z$, $\hbar\rightarrow 1$, $D\rightarrow 1/\left( 2n_0 k_0\right)$, and $-V\rightarrow k_0\Delta n^2(x,z)/\left(2n_0\right)=k_0 \left[n^2(x,z)-n_0^2\right]/(2n_0)\approx k_0\Delta n$ are made. Above $k_0$ is the vacuum wavenumber and $n_0$ the unmodified refractive index of the material. The value $D=1.27\times 10^{-14}~$Jm$^2$ used in the numerical simulations thus corresponds to a vacuum wavelength of $1~\mu$m, $l=1~\mu$m, a refractive index profile $\Delta n$ equal to the potential $V$ provided in the text, and a normalized propagation length $t_0\approx 12.5\mu$m. Thus, when the modulation period $T/t_0=\Lambda=0.8$ corresponds to a longitudinal spatial modulation of 10~$\mu$m. % with finally $\mu$s (ms) corresponding to $\mu$m (mm). 

\section*{acknowledgement}
 C.P.J. has received funding from the European Union’s Framework Programme for Research and Innovation Horizon 2020 under the Marie Sklowdowska-Curie Grant Agreement No. 889525. This research has been supported by Deutsche Forschungsgemeinschaft (DFG) via the International Research Training Group GRK 2101.  This work is supported by the DFG Collaborative Research Center "NOA – Nonlinear Optics down to Atomic scales", Grant No. SFB 1375. %A.A. acknowledges support by Deutsche Forschungsgemeinschaft (DFG) via the International Research Training Group GRK 2101. We acknowledge the Max Planck School of Photonics supported by BMBF, Max Planck Society, and Fraunhofer Society

\bibliography{references}

\end{document}